\def\be{\begin{equation}}
\def\ee{\end{equation}}
\def\ber{\begin{eqnarray}}
\def\eer{\end{eqnarray}}
\def\bers{\begin{eqnarray*}}
\def\eers{\end{eqnarray*}}

\def\JPCM{J. Phys.: Condens. Matter}
\def\PR{{ Phys. Rev.}\ }
\def\PRL{{ Phys. Rev. Lett.}\ }

\def\JPCM{{ J. Phys.: Condens. Matter}\  }

\documentclass[aps,prb,twocolumn,groupedaddress,showpacs,amsmath,amssymb]{revtex4-1}
\usepackage{dcolumn}   % Align table columns on decimal point
\usepackage{amsmath}
\usepackage{graphicx}    % Include figure files
\usepackage{subfigure}
\usepackage{color}
\newcommand{\comment}[1]{}
\usepackage{ifthen}
\newboolean{includefigs}
\setboolean{includefigs}{true}      % Permits rapid compiling by removing figures/text
\newboolean{includetext}
\setboolean{includetext}{true}     % false = removed, true = included
\newcommand{\condcomment}[2]{\ifthenelse{#1}{#2}{}}
\begin{document}

%%%%%%%%%%%%%%%%%%%%%%%%%%%%%%%%%%%%%%%%%%%%%%%%%
%               TITLE 
%%%%%%%%%%%%%%%%%%%%%%%%%%%%%%%%%%%%%%%%%%%%%%%%%
\title{Chemically-Mediated Quantum Criticality in NbFe$_2$}

\author{Aftab Alam$^{1}$ and D. D. Johnson$^{1,2}$}
\email[emails: ]{ddj@ameslab.gov,aftab@ameslab.gov}
\affiliation{$^{1}$Division of Materials Science and Engineering, Ames Laboratory, Ames, Iowa 50011;}
\affiliation{$^{2}$Department of Materials Science \& Engineering, Iowa State University, Ames, Iowa 50011.}

%%%%%%%%%%%%%%%%%%%%%%%%%%%%%%%%%%%%%%%%%%%%%%%%%
%               ABSTRACT 
%%%%%%%%%%%%%%%%%%%%%%%%%%%%%%%%%%%%%%%%%%%%%%%%%
\begin{abstract}
Laves-phase Nb$_{1+c}$Fe$_{2-c}$ is a rare itinerant intermetallic compound exhibiting magnetic quantum criticality at $c_{cr}$$\sim$$1.5\%$Nb excess; its origin, and how alloying mediates it, remains an enigma.
For NbFe$_2$, we show that an unconventional band critical point (uBCP) above the Fermi level E$_\text{F}$ explains most observations, and that chemical alloying mediates access to this uBCP by an increase in E$_\text{F}$ with decreasing  electrons (increasing \%Nb), counter to rigid-band concepts.
We calculate that E$_\text{F}$ enters the uBCP region for $c_{cr}>1.5\%$Nb and by $1.74\%$Nb there is no Nb site-occupation preference between symmetry-distinct Fe sites, i.e., no electron-hopping disorder, making resistivity near constant as observed.
At larger Nb (Fe) excess, the ferromagnetic Stoner criterion is satisfied. 
\end{abstract} 
\date{\today}
\pacs{71.20.Lp,75.10.Lp,75.45.+j}
\maketitle

%%%%%%%%%%%%%%%%%%%%%%%%%%%%%%%%%%%%%%%%%%%%%%%%%
%                INTRODUCTION
%%%%%%%%%%%%%%%%%%%%%%%%%%%%%%%%%%%%%%%%%%%%%%%%%

{\par} Quantum criticality emerges from collective low-energy excitations leading to a second-order phase transition at zero temperature (T).\cite{Sachdev2000} 
In the study of correlated-electron materials, e.g., high-T superconductors and heavy-Fermion compounds, understanding such critical phenomenon remains a principal challenge; and, locating any existing quantum critical points (QCPs) is difficult. 
Near these QCPs, quantum fluctuations (rather than thermal fluctuations) are observed to give rise to exotic effects, e.g.,  non-Fermi liquid behavior.\cite{Lohneysen2007}  
In intermetallic compounds the situation appears simpler, where, by varying non-thermal order parameters, such as applied pressure, external magnetic field, or chemical doping, various phases near a QCP can be accessed, i.e., ferromagnetic (FM), antiferromagnetic (AFM), and paramagnetic (PM) states. 
Importantly, strong electron-electron interactions are not required for an alloy to show correlated behavior, as now confirmed in Fe-As compounds.\cite{McQueeney2011} 
The Laves phase of Nb$_{1+c}$Fe$_{2-c}$ is one important example and featured in recent reviews of quantum criticality in weak magnets.\cite{Schofield2010}
At stoichiometry the susceptibility exhibits Curie-Weiss behavior down to a spin-density wave (SDW) transition at T$_{\text{sdw}}\simeq 10$~K, whereas at $\sim$$1.5\%$Nb excess a QCP is observed where the SDW collapses and non-Fermi-liquid behavior occurs.\cite{Brando,Moroni-Klementowicz}
For larger Nb \emph{excess} ($c>0$, hole doping) or Fe \emph{excess} ($c<0$, electron doping) a FM transition is always observed.\cite{Brando,Moroni-Klementowicz}
Indeed, the sensitivity of the magnetic state to Nb deficiency has long been known.\cite{Shiga,Yamada} 

{\par} Although well characterized experimentally, our understanding of the properties of Nb$_{1+c}$Fe$_{2-c}$ is lacking, especially how chemical disorder affects the magnetic transitions and QCP, and the FM onset at larger dopings.
Recent theoretical work studied the electronic properties of NbFe$_2$,\cite{DJSingh2010} but did not address the critical chemical effects.
Of course, including properly the effects of disorder in these class of systems, specifically at such a small doping ($c$), is a considerable challenge.
Attempting to address doping, recent studies use the virtual crystal approximation\cite{DJSingh2008} (VCA) or supercells (ordered array of dopants),\cite{Tompsett2010} having severe shortcomings, as we discuss.

{\par} In metals low-energy excitations lie at/near the Fermi surface (FS), and, thus, some unique spectral feature at/near E$_\text{F}$ is typically required to drive a transition, as with FS nesting\cite{Moss,Gyorffy} or van Hove band critical points\cite{vanHove} (BCP) types of ordering.\cite{Clark,Althoff}
Such FS features have been rarely identified as the origin of quantum criticality.
We detail how an unconventional BCP (uBCP), i.e., an accidental saddle-point dispersion, above E$_\text{F}$ is responsible for the QC behavior in metallic Nb$_{1+c}$Fe$_{2-c}$.
We show that chemical disorder mediates, via $d$-state hybridization, access to this uBCP even with fewer electrons for increasing \%Nb, which rigid-band/VCA cannot describe, effects well known in alloy theory.

{\par}  We use an all-electron, Korringa-Kohn-Rostoker (Green's function) electronic-structure method and the Coherent-Potential Approximation\cite{Johnson} (KKR-CPA)  to calculate the electronic dispersion, density of states, total energies and doping site-preferences in Nb$_{1+c}$Fe$_{2-c}$.
See Ref. \onlinecite{Alam2009_2010} for details and examples for (dis)ordered alloys.
In agreement with experiment, we calculated $c_{cr}$  ``onset'' at $1.5\%$Nb, and,  at $1.74\%$Nb, when E$_\text{F}$ lies exactly at the uBCP, we find that  Nb has no site-occupation preference between the two symmetry-distinct Fe sites, favoring a homogeneous solute distribution and no electron-hopping disorder, suggesting a near constant resistivity.
We find competing FM and AFM states, as observed, that are associated with the competing wavevectors from the uBCP.
We show that the FM Stoner criterion in Nb$_{1+c}$Fe$_{2-c}$ is obeyed for larger Nb (or Fe) excess.
If the uBCP are removed from consideration, none of these results hold.
We conclude that the QCP occurs via chemical-disorder-mediated access to the uBCP above E$_\text{F}$.

\begin{figure}[t]
\centering
\includegraphics[width=8.3cm]{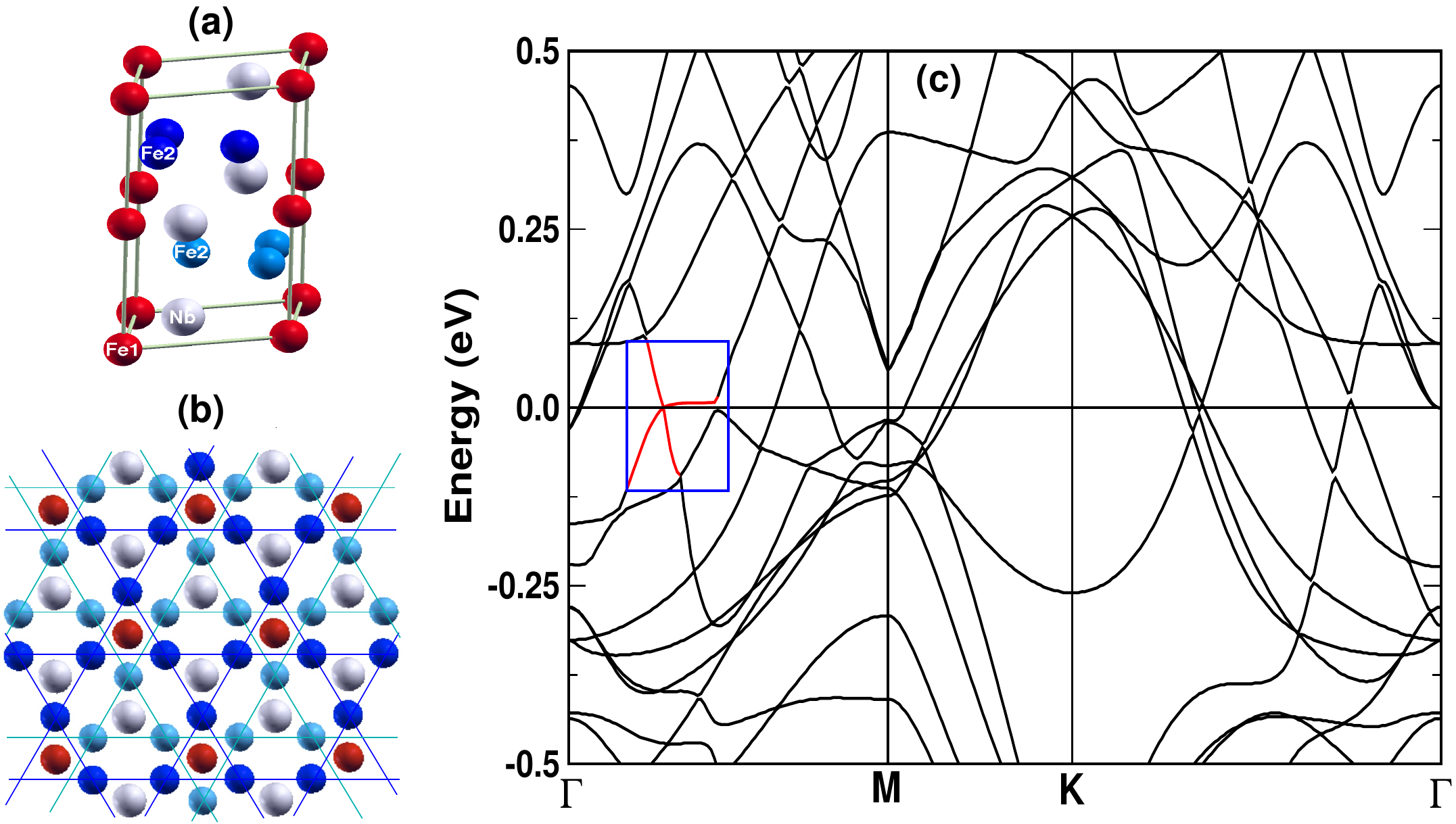}
\caption {(Color online) (a) Laves unit cell. (b) Fe$^{(6h)}$ sites form two Kagome nets, with Nb (gray), Fe$^{(2a)}$ (red) and Fe$^{(6h)}$ (blue), with higher (lower) planes shaded. (c) NbFe$_2$ bands (E$_{\text{F}}$ at $0$~eV). Highlighted box shows an uBCP above E$_{\text{F}}$.} 
\label{Fig_structure}
\end{figure}

{\par} To understand chemical disorder effects, we first need to appreciate the structure of NbFe$_2$, which crystallizes in a $C14$ hexagonal Laves phase with space group P$6_3/mmc~(\#194)$. 
In terms of crystallography, NbFe$_2$ $\equiv$ Nb$_4$Fe$_{2}^{(2a)}$Fe$_{6}^{(6h)}$ with a 12-atom unit cell: Fe$^{(6h)}$ sites form two Kagome networks ($\perp$ to $c$-axis) separated by Fe$^{(2a)}$ sites in a hexagonal sublattice, and Nb atoms occupy the interstices, see Fig.~\ref{Fig_structure}(a,b). 
The  Wykoff positions (without inversion) are: Nb at $4$f($\frac{1}{3},\frac{2}{3},x$), 
Fe$^{(2a)}$ at $2$a($0,0,0$) and Fe$^{(6h)}$ at $6$h($y,2y,\frac{3}{4}$). 
Calculations are performed with measured\cite{Brando} structural parameters $a=4.8401$ \r{A}, $c=7.8963$ \r{A} and internal coordinates $x=0.0629$ and $y=0.1697$. 
Using KKR-CPA we may study the effect of (in)homogeneous solute distributions.
Disorder with antisite Nb on one or both Fe sublattices is written as Nb$_4$(Fe$^{(2a)}_{1-c^{(2a)}}$Nb$^{(2a)}_{c^{(2a)}}$)$_2$(Fe$^{(6h)}_{1-c^{(6h)}}$Nb$^{(6h)}_{c^{(6h)}}$)$_6$, or Nb$_4$(Fe$_{1-c}$Nb$_{c}$)$_8$ for the homogeneous case.
 
{\par} To identify critical electronic features and chemical disorder effects, we detail the dispersion $\epsilon(\bf{k};\text{E})$ and density of states (DOS).
KKR uses constant-E matrix inversion to get $\epsilon(\{\bf{k}\};\text{E})$, rather than constant-{\bf k} diagonalization to get eigenvalues $\epsilon(\bf{k};\{\text{E}\})$.
To handle (dis)ordered cases, we calculate the Bloch spectral function\cite{Gyorffy} A$_B({\bf k};\text{E})$ on a grid of $32\times32\times24$ {\bf k}-points to project the dispersion. 
For ordered cases, A$_B({\bf k};\text{E})$ yields $\delta(\text{E} - \epsilon(\bf{k};\text{E}))$, i.e., bands; otherwise it exhibits disorder-induced spectral broadening in E and $\bf{k}$, related to the finite electron scattering length.   

%%%%%%%%%%%
% FIGURE 1: Structure and Dispersion

\begin{figure}[t]
\centering
\includegraphics[width=8cm]{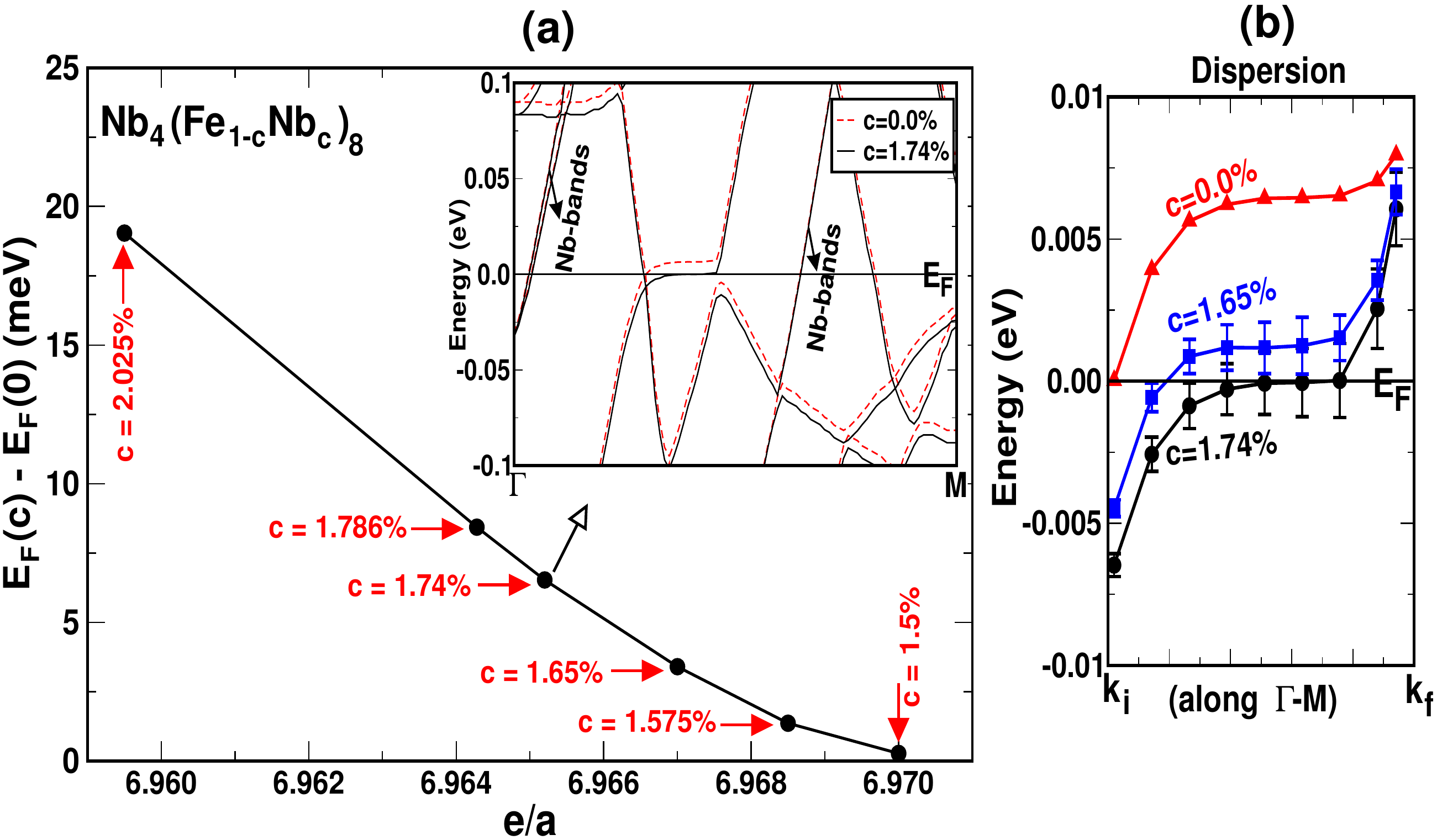}
\caption{(Color online) (a) $\Delta$E$_{\text{F}}(c)$ shift due to \%Nb-excess vs. $e/a$ for Nb$_4$(Fe$_{1-c}$Nb$_{c}$)$_8$. E$_{\text{F}}$ lies at uBCPs at $c_{cr}=1.74\%$ but enters spectral tails at $1.65\%$. 
Inset (a) shows (un)doped A$^{max}_B({\bf k};\text{E})$ along $\Gamma$-M, and (b) expands around uBCPs (bars are spectral widths due to disorder broadening). }
\label{doping_a+h}
\end{figure}

{\par} We now show that the $c$-dependence tied to the observed quantum criticality arises from the NbFe$_2$ dispersion above E$_{\text{F}}$.
Our bands in Fig.~\ref{Fig_structure}(c) agree with those from full-potential methods.\cite{DJSingh2010} 
The bands crossing E$_{\text{F}}$ along $\Gamma$-M arise mainly from Fe$^{(6h)}$ t$_{2g}$-orbitals, and lead to \emph{saddle-point dispersion} slightly ($6.6$ meV) above E$_{\text{F}}$ with an unusual flat dispersive region (uBCP), Figs.~\ref{Fig_structure}(c) and~\ref{doping_a+h}(b). These uBCP near E$_{\text{F}}$ are not a result of symmetry, but arise from accidental band crossings. 
Experimentally, Crook and Cywinski\cite{Crook94} inferred that the Fe$^{(6h)}$  t$_{2g}$-orbitals in the Kagome nets plays a critical role in the competing magnetic order associated with the quantum phase transition (QPT).
The Nb$_{1+c}$Fe$_{2-c}$ phase diagram\cite{Brando} shows the QCP onset at ambient pressure at $c_{cr}\sim1.5\%$Nb (hole-doped) and extends to $c\simeq2.0\%$, with FM at larger doping.
For a metal, only low-energy excitations near E$_{\text{F}}$ can be relevant for such low-T transitions.
By small tuning of a non-thermal order parameter, i.e., $c$, the unusual QPT behavior is observed, and can be explained if these uBCPs are accessed.

{\par}For homogeneous doping, we find a chemical-disorder-mediated increase of E$_{\text{F}}(c)$ versus \%Nb excess, or decreasing electron-per-atom ($e/a$) ratio, Fig.~\ref{doping_a+h}(a).
Counter to rigid-band concepts, E$_{\text{F}}$ rises to uBCP and due to Fe-Nb (\emph{bond/antibond}) alloying hybridization in Kagome nets Fe-bands shift lower but bands from the pure Nb-layer remain unaffected, Fig.~\ref{doping_a+h}(a) inset.
(An ordered array of impurities exaggerates the effect, see below, showing disorder plays a key role.) 
At $1.74\%$Nb excess, uBCP lay at E$_{\text{F}}$, a $6.6~m$eV shift due to alloying and disorder (finite life-time) effects, see Fig.~\ref{doping_a+h}(b).
The $6.6~m$eV (or $77$~K) sets the maximum temperature, as observed, for these effects to occur on stoichiometry. 
The dispersion and disorder-induced widths along $\Gamma-$M for $0 \le c \le c^{\text{cpa}}_{cr}$, Fig.~\ref{doping_a+h}(b), estimates the QC range.
%By $1.60\%$Nb, E$_{\text{F}}$ is within $2~m$eV of the uBCP; 
By $1.65\%$, E$_{\text{F}}$ enters the spectral tails, giving zero-energy excitations into the anomalous dispersion; E$_{\text{F}}$ is maximally aligned with the uBCP by $1.74\%$ (giving a Lifshitz-type transition, see below), and exits by $2\%$, where FM is observed.
We conclude that the QCP occurs from alloying/disorder-mediated access to the uBCP inherent in NbFe$_2$ dispersion  above E$_\text{F}$, detailed more below. 
     
%Effect of Doping (at both 2a and 6h-sites) on E_F 

%%%%%%%%  SUPERCELLS
{\par} We have performed supercell calculations to illustrate how sensitive $\Delta$E$_{\text{F}}(c)$ is to approximations used to address chemical effects, which ignore disorder. 
Supercells are numerically costly due to large cells needed with decreasing \%Nb. 
We performed calculations at two concentrations with cells constructed by substituting a Nb-atom for one Fe-atom in a $2\times2\times1$ ($6.25\%$Nb excess) and a $2\times 2\times 2$ ($3.125\%$Nb excess) supercell. 
The supercells also yield a $\Delta$E$_{\text{F}}(c)$ increase, relative to $c=0$\%, of $117.1$ and $281.3~m$eV for $3.125\%$ and $6.25\%$Nb excess, respectively.
The CPA shifts are $65.4$ and $158.1~m$eV, respectively, which lie on the curve in Fig. \ref{doping_a+h}(a) at smaller $e/a$. 
So, supercells do not provide the correct occupancy probability nor hybridization across the Kagome net, missing the key disorder effects.
Nonetheless, supercell results do reinforce the fact that rigid-band/VCA concepts are invalid, missing the critical alloying and disorder effects.

%Compare stoich DOS with off-stoich DOS at c=1.74 
\begin{figure}[t]
\centering
\includegraphics[width=7cm]{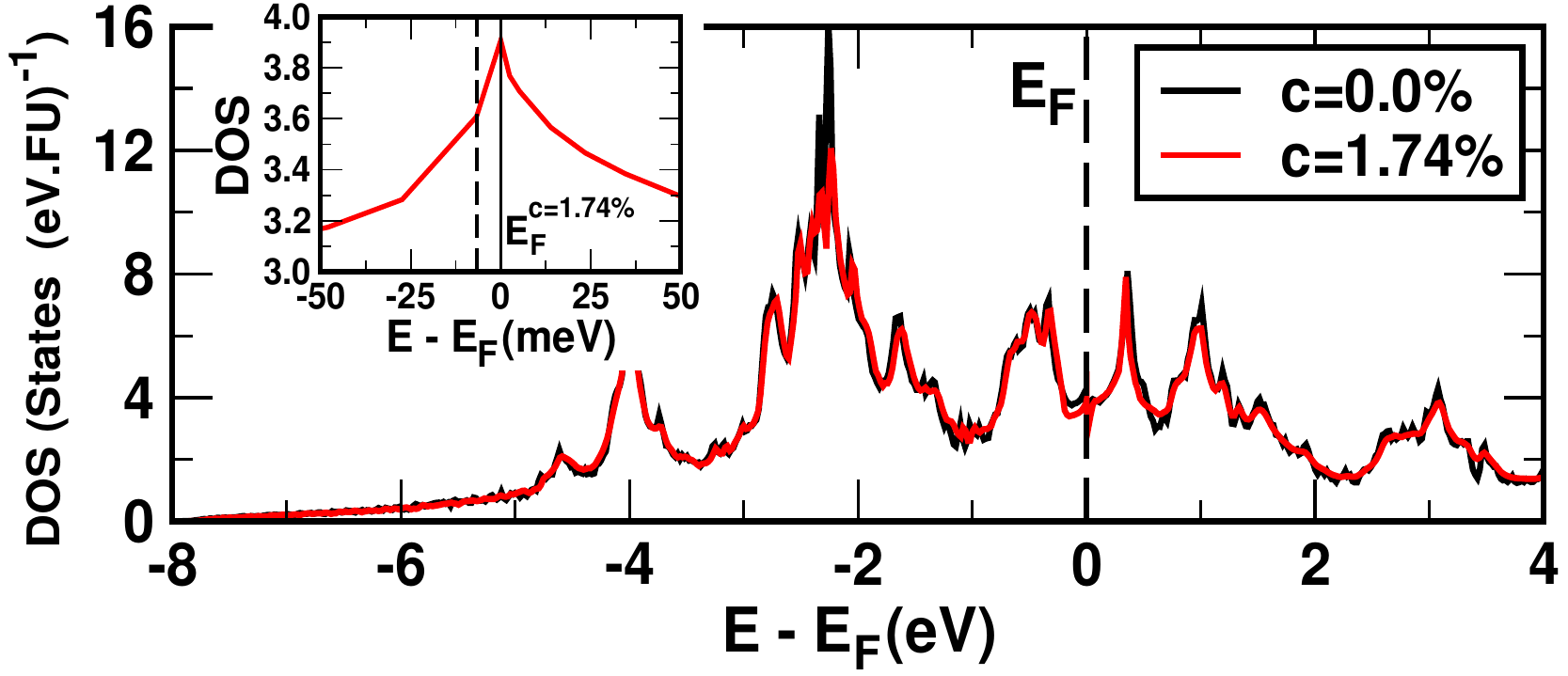}
\caption
{(Color online) Non-magnetic DOS per formula unit relative to E$_{\text{F}}$ for Nb$_{1+c}$Fe$_{2-c}$ at c$_{cr}=0$ and $1.74\%$. Inset: DOS at $c_{cr}$ around E$_{\text{F}}$ (vertical dashed line is E$_{\text{F}}$ at c=$0\%$).  }
\label{dos1}
\end{figure}

{\par}Figure~\ref{dos1} compares the DOS for Nb$_{1+c}$Fe$_{2-c}$ at $c_{cr}$ to $0\%$, which is similar to full-potential results.\cite{DJSingh2010,Tompsett2010} 
Disorder broadening for $c>0\%$ is evident. 
E$_{\text{F}}$ lies near a precipice of a DOS depression, which plays a role in forming a SDW state at $c=0\%$. 
$n(\text{E}_{\text{F}})$ is $3.59$~states~(eV - formula~unit)$^{-1}$ for NbFe$_2$. 
Notably, for NbFe$_2$, unlike conventional BCP,\cite{vanHove} the {\it {saddle-point dispersion}} at ${\bf k}$'s associated with the uBCP is (beyond) cubic ($\epsilon_k \propto k_{x}^{3}$) in one direction and quadratic in the orthogonal $k_y$,$k_z$ plane, yielding an chemically-mediated peak in $n$(E$_{\text{F}}$) when E$_{\text{F}}$ and uBCP are aligned, Fig.~\ref{dos1} inset.  
From Stoner theory with interaction parameter $I$,\cite{Stoner1939} a FM instability occurs if  $n(\text{E}_{\text{F}}) I>1$. 
For pure Fe $d$-electrons, $I$ was reported\cite{Andersen1977} between $0.7-0.9~$eV.
From susceptibility data for NbFe$_2$, the Stoner factor $[1 - n(\text{E}_{\text{F}}) I ]^{-1}$  was estimated\cite{Brando} at $\sim$$100$. 
Our calculated $I$ for Nb$_{1+c}$Fe$_{2-c}$ is almost constant versus $c$ ($0.88\le I \le 0.9$), but $n(\text{E}_{\text{F}})$ increases with doping, increasing $n(\text{E}_{\text{F}}) I$, e.g.,
$n(\text{E}_{\text{F}})$ for $1.74\%$Nb increases to $3.90$  states-$(\text{eV\ FU})^{-1}$.
Stoner's criterion is satisfied beyond $2\%$Nb, as observed, and discussed more below.

{\par}Lifshitz transitions are mediated by FS topology changes (e.g., collapse of FS neck or loss of pockets). 
An unconventional Lifshitz transition emerging near a {\it marginal} QCP was proposed in ZrZn$_2$.\cite{Lifshitz} 
Such a zero-temperature, pressure-induced  transition is also associated with access to an uBCP by an increase in E$_{\text{F}}$.
The FS topology changes are reflected in a maximum in $n(\text{E}_{\text{F}})$ at the Lifshitz point, as we found for hole-doping in NbFe$_2$ at c$_{\text{cr}}=1.74\%$ (Fig.~\ref{dos1} inset). 
Not only the topology, but the FS volume is strongly dependent on doping, and in NbFe$_2$ enhanced with hole doping, and also observed in a hole-doped Ba$_{0.3}$K$_{0.7}$Fe$_{2}$As$_2$.\cite{Nakayama}  
   
\begin{figure}[b]
\centering
\includegraphics[width=5.25cm]{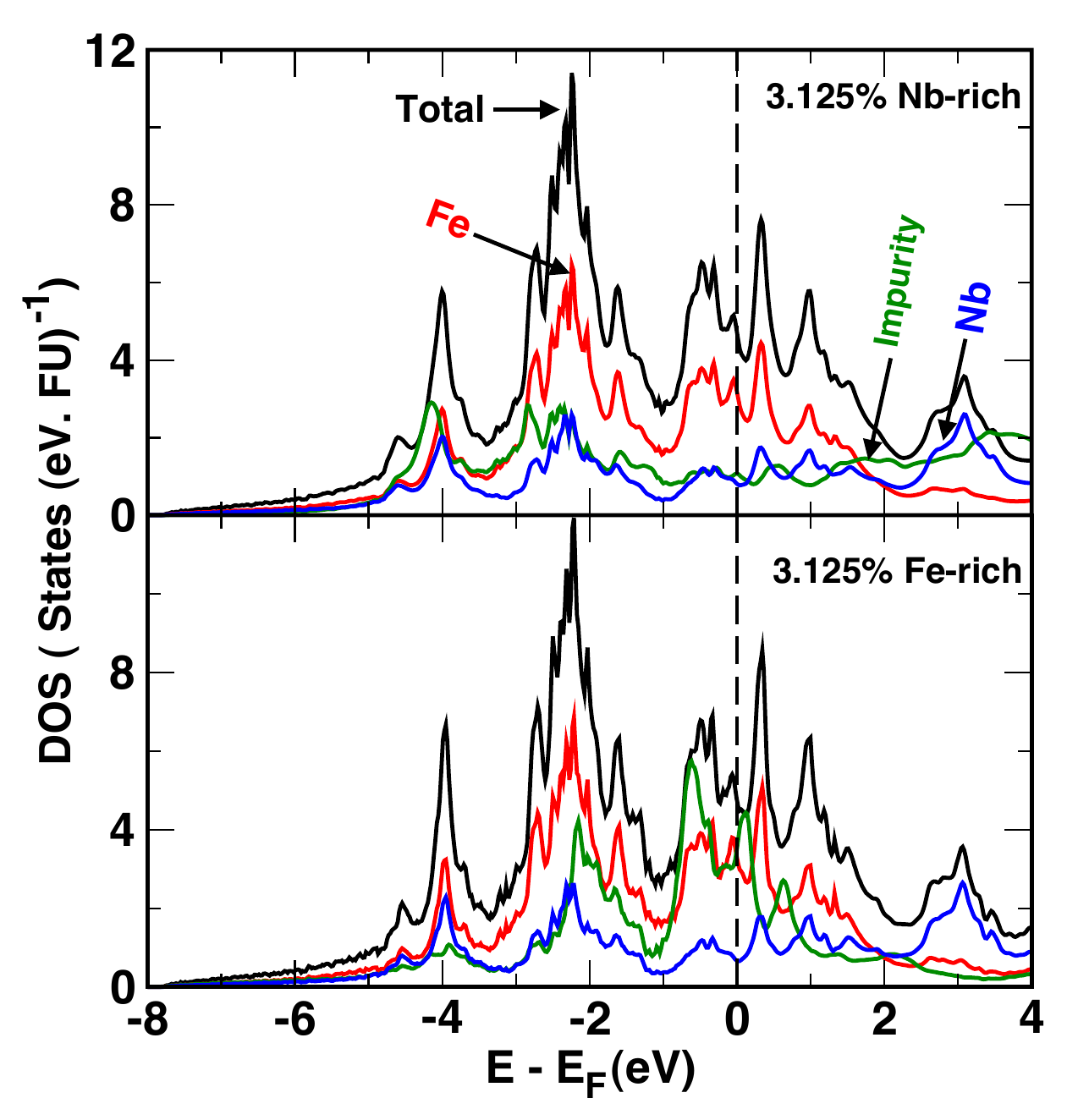}
\caption{(Color online) Non-magnetic DOS relative to E$_\text{F}$ for $3.125\%$ Nb-rich (top) and
 $3.125\%$ Fe-rich (bottom). }
\label{dos2}
\end{figure}

{\par}To explore beyond the QCP, we studied doping in the Fe- and Nb-rich parts of the phase diagram. 
The DOS (atom- and impurity-projected) for $3.125\%$ Nb- and Fe-rich Nb$_{1+c}$Fe$_{2-c}$ is shown in Fig.~\ref{dos2}. 
The Nb-rich (hole-doped) $n(\text{E}_{\text{F}})$ increases to $4.66$ states-$(\text{eV\ FU})^{-1}$. 
Fe-rich (electron-doped) alloys behave opposite to what is expected from rigid-band theory; i.e. with electron doping, $n(\text{E}_{\text{F}})$ rises to $4.75$ states-$(\text{eV\ FU})^{-1}$, which is mainly due to Fe-impurity DOS originating from Nb-layers, see Fig.~\ref{dos2}~(lower panel). 
These values of Nb-rich and Fe-rich $n(\text{E}_{\text{F}}$) satisfy the FM Stoner criterion, as observed.\cite{Brando}

% DOS at 3.125% Nb-rich and 3.125% Fe-rich 

{\par}We have also studied how the nature of the uBCP (or QCP) is affected by inhomogeneous Nb doping on the two Fe sublattices. 
We find that, if only Fe$^{(6h)}$ sites are doped, the uBCP lies at E$_{\text{F}}$ when c$_{cr}^{(6h)}=2.34\%$, for an average c$_{cr}^{avg}=2.34\% \times (6/8)=1.75\%$.
For Fe$^{(6h)}$-only case, dispersion is similar to the homogeneous case as in Fig.~\ref{Fig_structure}(c), because mainly Fe$^{(6h)}$ t$_{2g}$-orbitals within the Kagome sheets are involved.
In contrast, if only Fe$^{(2a)}$-sites are doped, the uBCP lies at E$_{\text{F}}$ when c$_{cr}^{(2a)}=8.7\%$, i.e., c$_{cr}^{avg}=8.7\% \times (2/8)=2.02\%$. 
The shift of E$_{\text{F}}$ relative to dispersion in this case arises due to hybridization of Fe$^{(2a)}$-states indirectly with Fe$^{(6h)}$-states.
Dispersion with Fe$^{(2a)}$-only doping shows an increased slope of the uBCP compared to the other cases, changing the character and $c$-dependence of the response.

\begin{figure}[t]
\centering
\includegraphics[width=6.0cm]{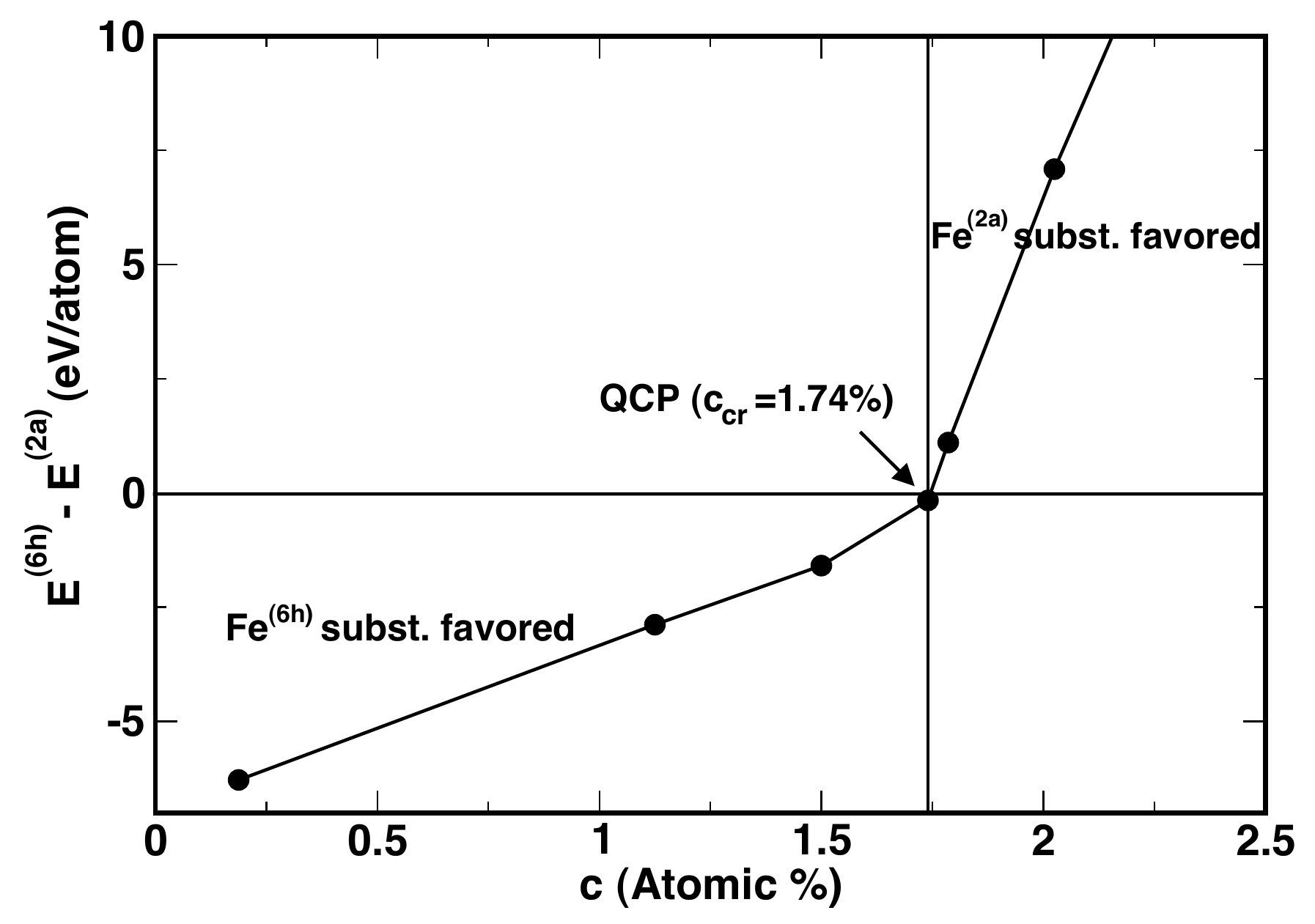}
\caption {Nb site-preference energy difference (E$^{(6h)}-$E$^{(2a)}$)  versus \%Nb excess in Nb$_4$(Fe$_{1-c}$Nb$_{c}$)$_8$. 
E$^{(\alpha)}$ is the total energy when only the Fe$^{(\alpha)}$-sites are doped.
At $c_{cr}=1.74\%$, no Fe-site is favored, i.e., homogeneous solute distribution.}
\label{site_pref_E}
\end{figure}

{\par}In Fig.~\ref{site_pref_E} we report Nb site-preference energy differences (E$^{(6h)}-$E$^{(2a)}$) versus \%Nb-excess in Nb$_4$(Fe$^{(2a)}_{1-c^{(2a)}}$Nb$^{(2a)}_{c^{(2a)}}$)$_2$(Fe$^{(6h)}_{1-c^{(6h)}}$Nb$^{(6h)}_{c^{(6h)}}$)$_6$, where E$^{(\alpha)}$ is the total energy when only the Fe$^{(\alpha)}$-sites are doped.
At $c^{cpa}_{cr}=1.74\%$, where the uBCP lies at E$_{\text{F}}$, there is no energetically favored site occupancy, therefore, no electron-hopping disorder  effects, explaining the observed extremely slow variation of $\rho$ versus doping.\cite{Brando}
With no site preference, a homogeneous solute distribution is favored, further supporting our focus on this case.
These results shed light on the microscopic phenomenon occurring at/near the QCP: the alignment of E$_{\text{F}}$ with the uBCP at $1.74\%$ provides maximum response concomitant with no electron-hopping disorder and instability to both AFM and FM fluctuations.

%concentration dependence of site-preferential doping

{\par} We now explore the stability of competing magnetic phases. 
Out of several small-cell, magnetic configurations in NbFe$_2$, we found a AFM state energetically favored by $-10.45~m$eV/atom over the non-magnetic state, where local moments on the Fe$^{(2a)}$- and Fe$^{(6h)}$-sites are aligned antiparallel with values of $0.69\ \mu$B and  $-0.98\ \mu$B, respectively, and the Nb has an induced moment of $0.14\ \mu$B aligned with the Fe$^{(2a)}$-site. 
The same ground state was found in other calculations.\cite{DJSingh2010,Tompsett2010} 
The phase diagram indicates a SDW state for NbFe$_2$ with a highly itinerant nature (as for iron-arsenides\cite{McQueeney2011}), which competes with nearby FM states. 
The itineracy of magnetism is clear from the spin density on Fe-sites, see Supplement, while carrier density peaks at $c_{cr}$, Fig.~\ref{dos1} inset.

{\par} To connect to scattering experiments, the uBCP $\epsilon(\bf{k})$ exhibit zero-velocity quasiparticles near E$_{\text{F}}$ at symmetry-equivalent, non-special {\bf k}-points $Q_{0}(\pm1,0)$ and $Q_{0}(\pm\frac{1}{2},\pm\frac{\sqrt3}{2})$, where $Q_{0}\approx0.25$ in units of the $\Gamma-$M caliper in Fig.~\ref{Fig_structure}(c).
These $Q$'s remain relevant for off-stoichiometric alloys.
The susceptibility, i.e.,
\be
\chi(\mathbf{q}) = \sum\nolimits_{\mathbf{k}}
             [f(\epsilon_{\mathbf{k}}) - f(\epsilon_{\mathbf{k}+ \mathbf{q}})]
             [\epsilon_{\mathbf{k}+ \mathbf{q}} -  \epsilon_{\mathbf{k}}]^{-1}  \ \ ,
\ee
can exhibit an enhanced response due to a convolution of (un)occupied states near E$_F$, as for FS nesting.\cite{Moss,Gyorffy}
Albeit weaker, it can occur from saddle-point topology too.
From the small caliper of the flat part of the uBCP, i.e., $0.055 (2\pi/a)$, we estimate a $|q|$ of $0.07~\r{A}^{-1}$ close to the $0.05~\r{A}^{-1}$ estimated from experiment.\cite{Moroni-Klementowicz} 

{\par} Thus, from our direct calculations or estimated $\chi(\mathbf{q})$ features, PM, FM and SDW (AFM) states compete near the QCP, until overwhelmed by a Stoner instability at larger $|c|$, shown above.
 Such competing magnetic behavior is observed\cite{Moroni-Klementowicz} near the QCP, where  the low-T resistivity ($\rho(T)\sim T^{\nu}$) exponent $\nu$ varies between $3/2$ and $5/3$, giving unusual non-Fermi liquid behavior.

{\par}Quantifying the anomalous response further requires $\chi(\mathbf{q},c,\omega)$, which is beyond the present scope. But, the change in $\omega$=0 susceptibility, $\Delta \chi(Q,\omega$=$0)$, yields the change in the DOS at E$_{\text{F}}$, i.e., $\Delta n(\text{E}_{\text{F}})$, and, if large, a FM instability.
Neal \emph{et al.} have extended the Moriya $\chi(\mathbf{q},c=0,\omega)$ theory to account for uBCP, which yields an anomalous frequency response and competing FM and SDW states,\cite{Pickett2010} agreeing with our direct calculations.

{\par}  In summary, we have identified an accidental Fermi-surface (non-ideal saddle-point) dispersion as the origin for observed behavior associated with the quantum criticality in Nb$_{1+c}$Fe$_{2-c}$.
We find that Nb (hole) doping accesses an unconventional band critical points in NbFe$_{2}$ that provide the necessary low-energy excitations for a (Lifshitz-type) transitions.
This origin explains most of the observed doping behavior in this QC intermetallic compound, specifically,
($i$) onset of $c_{cr}$ \%Nb-excess for the QCP; 
($ii$) almost constant resistivity versus $c$, 
($iii$) competing magnetic states and temperature scale,  
($iv$) observed SDW wavevectors, and, finally,
 ($v$) the stable FM states at large hole- or electron-doping ($>$2\%).
To explore the nature of the QC transition further, these electronic and chemical features can be incorporated into a model Hamiltonian, but they would not have been discovered without the full dispersion and chemical alloying effects detailed here.

%%%%%%%%%%%%%%%%%%%%%%%%%%%%%%%%%%%%%%%%%%%%%%%%%
%                     Acknowledgements
%%%%%%%%%%%%%%%%%%%%%%%%%%%%%%%%%%%%%%%%%%%%%%%%%
\vspace{0.1cm}
Work funded by the U.S. Department of Energy BES-DMSE (DE-FG02-03ER46026) and Ames Laboratory (DE-AC02-07CH11358), operated by Iowa State University.
We thank to W. Pickett for suggesting this problem.
%%%%%%%%%%%%%%%%%%%%%%%%%%%%%%%%%%%%%%%%%%%%%%%%%%
%                     Reference Page
%%%%%%%%%%%%%%%%%%%%%%%%%%%%%%%%%%%%%%%%%%%%%%%%%%
\vspace{-0.5cm}
%\bibliography{Aftab}

\end{document}